\documentclass{emulateapj}
\usepackage{apjfonts}

\shorttitle{{\it Suzaku} observation of Mrk~421}
\shortauthors{M. Ushio et al.}

\begin{document}
\title{Suzaku Wide Band Analysis of the X-ray Variability of 
TeV Blazar Mrk~421 in 2006}


\author{Masayoshi Ushio\altaffilmark{1,2}, Takaaki Tanaka\altaffilmark{3}, 
Grzegorz Madejski\altaffilmark{3}, Tadayuki Takahashi\altaffilmark{1,2}, 
Masaaki Hayashida\altaffilmark{3}, \\Jun Kataoka\altaffilmark{4},
Daniel Mazin\altaffilmark{5}, Stefan R\"ugamer\altaffilmark{6}, 
Rie Sato\altaffilmark{1}, Masahiro Teshima\altaffilmark{7}, 
Stefan Wagner\altaffilmark{8} and Yuichi Yaji\altaffilmark{9}}

\email{ushio@astro.isas.jaxa.jp}


\altaffiltext{1}{Department of High Energy Astrophysics, 
Institute of Space and Astronautical Science (ISAS), 
Japan Aerospace Exploration Agency (JAXA), 3-1-1 Yoshinodai, 
Sagamihara, 229-8510}
\altaffiltext{2}{Department of Physics, University of Tokyo, 
Hongo 7-3-1, Bunkyo, 113-0033}
\altaffiltext{3}{Stanford Linear Accelerator Center, 
2575 Sand Hill Road, Menlo Park, CA 94025, USA}
\altaffiltext{4}{Department of Physics, Tokyo Institute of Technology, 
Ohokayama, Meguro, Tokyo, 152-8551, Japan}
\altaffiltext{5}{Institut de F\'isica d'Altes Energies, 
08193 Bellaterra, Spain}
\altaffiltext{6}{Universit\"at W\"urzburg Am Hubland 97074 
W\"urzburg, Germany}
\altaffiltext{7}{Max-Planck-Institut f\"ur Physik, D-80805 
M\"unchen,Germany}
\altaffiltext{8}{Landessternwarte, Universi\"at Heidelberg, 
K\"onigstuhl, 69117 Heidelberg, Germany}
\altaffiltext{9}{Saitama University, 255 Shimo-Okubo, Sakura, Saitama 338-8570}


\begin{abstract}
We present the results of X-ray observations of 
the well-studied TeV blazar Mrk~421 with the {\it Suzaku} satellite
in 2006 April 28.
During the observation, Mrk~421 was undergoing a large flare 
and the X-ray flux was variable, decreasing by $\sim 50$~\%, 
from $7.8\times 10^{-10}$ to $3.7\times 10^{-10}$ erg s${}^{-1}$cm${}^{-2}$ 
in about 6 hours, followed by an increase by $\sim 35$~\%.  
Thanks to the broad bandpass coupled with high-sensitivity of {\it Suzaku}, 
we measured the evolution of the spectrum over the 0.4--60~keV band 
in data segments as short as $\sim 1$ ksec.  
The data show deviations from a simple power law model, but also 
a clear spectral variability. 
The time-resolved spectra are fitted by a synchrotron model, where 
the observed spectrum is due to a exponentially cutoff power law 
distribution of electrons radiating in uniform magnetic field; 
this model is preferred over a broken power law.  
As another scenario, we separate the spectrum into ``steady'' 
and ``variable'' components by subtracting the spectrum 
in the lowest-flux period from those of other data segments.  
In this context, the difference (``variable'') spectra are all well described 
by a broken power law model with photon index $\Gamma \sim 1.6$, 
breaking at energy $\epsilon_{\rm brk}\simeq 3$~keV 
to another photon index $\Gamma \sim 2.1$ above the break energy, 
differing from each other only by normalization, while the spectrum of 
the ``steady'' component is best described by the synchrotron model.  
We suggest the rapidly variable component is due to relatively localized shock
(Fermi I) acceleration, while the slowly variable (``steady'') component is 
due to the superposition of shocks located at larger distance 
along the jet, or due to other acceleration process, such as the stochastic 
acceleration on magnetic turbulence (Fermi II) in the more extended region.

\end{abstract}

\keywords{acceleration of particles --- 
BL Lacertae objects: individual (Mrk~421) --- 
galaxies: jets --- X-rays: galaxies}

\section{Introduction}
Blazars, a sub-category of Active Galactic Nuclei (AGN), 
are characterized by broadband non-thermal emission 
and violent variability on time scales 
from a fraction of an hour to a few days with strong spectral evolution.  
This behavior is best described as 
their emission arising from Doppler-boosted relativistic jets 
which in turn dominates over the thermal signatures seen in other AGN
\citep[e.g.,][]{urry95, ulrich97}. 
The broadband spectra of blazars consist of two peaks, 
one in the radio to optical--UV range (and in some cases, 
reaching to the X-ray band), and the other 
in the hard X-ray to $\gamma$-ray region.
The high polarization of the radio to optical emission suggests 
that the lower energy peak is produced via the synchrotron process 
by relativistic electrons in the jet \citep[see, e.g.,][]{angel80}. 
The higher energy peak is believed to be due to Compton 
up-scattering of seed photons by the relativistic electrons.
A number of blazars with the synchrotron peak located 
in the X-ray range have also been detected in the TeV band, 
and the recent advances of TeV Cherenkov telescopes 
have revealed the existence of significant number of such 
TeV emitting blazars \citep[see][for recent synoptic study]{wag08}.
These objects provide the most direct evidence 
of efficient particle acceleration up to TeV energies, 
potentially providing the key information about the jet's 
composition and power, and thus the connection of it  
the central engine.

Mrk~421 is one of the nearest (z=0.031) and brightest 
TeV $\gamma$-ray emitting blazars: 
it was found as the first extragalactic TeV $\gamma$-ray 
emitting object \citep{punch92}, 
and has been repeatedly confirmed as a TeV source 
by various ground-based telescopes \citep[e.g.,][]{aha05b, albert07}.
It has also been one of the most extensively studied blazars, 
and has been a target of several multi-wavelength campaigns 
\citep[e.g.,][]{tad96, tad00, Tosti98, Rebillot06, fossati08, lichti08}.
Detailed studies during large amplitude flares are 
particularly important as they give us vital information
allowing studies of the underlying physics of the source.
In particular, the simultaneous observations 
in X-ray and TeV $\gamma$-ray band revealed a strong correlation 
between them and depict the picture 
in which the non-thermal distribution of relativistic electrons accelerated 
up to TeV energies are responsible for the variability of both bands.
Such a scenario - where lower energy component is synchrotron radiation 
from the relativistically accelerated electrons and the higher energy 
one is due to Compton up-scattering of the synchrotron photons 
by the same electron populations themselves - is widely adopted 
as the most plausible emission mechanism for TeV blazars, 
and is called one-zone Synchrotron Self-Compton (SSC) model 
\citep{inoue96,kataoka99}. 

Although the simplest model successfully explains the variability 
and the broadband spectra from radio to TeV $\gamma$ ray band to some extent,
recent improvements of atmospheric Cherenkov TeV $\gamma$-ray telescopes  
such as H.E.S.S, MAGIC and VERITAS make it possible 
to provide high quality spectra in relatively short exposures 
\citep[see, e.g.,][]{baix04,hint04,weekes02}.
Indeed, the recent observations with those telescopes 
imply the conventional model may no longer be applicable.
In particular, the ``orphan`` flares detected from 1ES~1959+650 
\citep{kraw04} and Mrk~421 \citep{blaz05},
during which the flux of TeV $\gamma$-ray flared up 
without counterparts in X-ray band, are difficult to explain 
within the context of simple one-zone SSC models.
Also, the detection of short variability time scale 
of TeV $\gamma$-ray emission from Mrk~421 
\citep[$\sim$ 10 minutes;][]{gaidos96} 
and PKS~2155-304 \citep[$\sim$ a few minutes;][]{aha07} 
may require unprecedentedly large Doppler factor 
$\delta \gtrsim 50$ \citep[e.g.,][]{ghisellini08},
which are much larger than ones consistently estimated 
from multi-wavelength analysis \citep[e.g.,][]{tave98,kubo98}.

In order to understand the particle acceleration 
and the non-thermal emission from TeV blazars, 
more sensitive observations are needed.
In particular, hard X-ray observations above 10~keV are 
important because the emission in this energy band 
reflects the behavior of the most energetic electrons 
and is a sensitive probe of their acceleration, cooling and escape rates.
In this paper, we present the results of the observations of Mrk~421 
with {\it Suzaku}, which is one of the most suitable observatories
for exploring such complex behaviors with its high sensitivity and 
wide-band coverage from the soft X-ray to hard X-ray band.
%
The observation was performed in 2006 April 28--29, 
at a time when the object was undergoing an outburst 
and showed a high level of activity.
In \S~2 and \S~3, we describe our {\it Suzaku} observations 
and the data reduction procedures. 
Analysis and results are shown in \S~4,  
and are discussed in the following \S~5, with summary in \S~6.
In the present work, we use the data products 
from the {\it Suzaku} pipeline processing version 2.0. 
The data reduction and analysis are done using {\tt HEADAS 6.3.1} 
and the spectral fitting is performed with {\tt XSPEC 11.3.2}.
Regarding the notation for both photon and electron energy,  
$\epsilon$ and $\nu$ are for photon energy and frequency, 
and $E$ and $\gamma$ are particle energy and its Lorentz factor, respectively.

\section{Observations}
The X-ray Observatory {\it Suzaku} \citep{mitsuda06}, 
developed jointly by Japan and  the US, 
has a scientific payload consisting of 
two kinds of co-aligned instruments; 
the XIS \citep{koyama06} and the HXD \citep{takahashi_hxd06,kokubun06}.
The XIS consists of four X-ray sensitive CCD cameras
which are located in the foci of X-ray telescopes \citep[XRT;][]{serlemit06}.
The HXD is a non-imaging detector system which
covers the hard X-ray bandpass of 10--600~keV with
PIN silicon diodes (10--60~keV) and GSO scintillators (40--600~keV).

\begin{figure} [htbp]
\begin{center}
\plotone{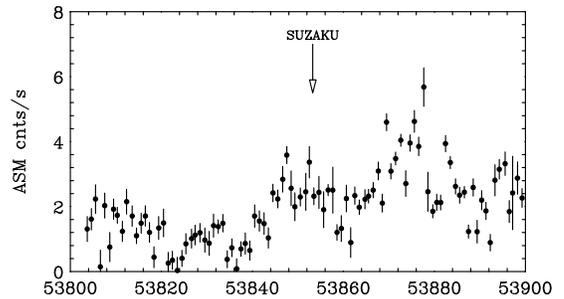}
\end{center} 
\caption{A long-term light curve of RXTE~ASM (2--10~keV) 
around the {\it Suzaku} observation.} 
\label{fig:lc_asm} 
\end{figure} 

We performed the {\it Suzaku} observation of Mrk~421 starting from
2006 April 28 (MJD 53853) 06:46 UT through April 29 (MJD 53854) 06:30 UT.
Figure~\ref{fig:lc_asm} shows a long term light curve of Mrk~421 
obtained with RXTE ASM \citep[2--10~keV, ][]{bradt93}, 
which includes the time interval of our {\it Suzaku} observation.
As shown in the figure, Mrk~421 exhibited high X-ray flux 
with a peak value reaching $\sim 2.5$ counts/sec around MJD = 53853, 
which is corresponding to $\sim$ 30~mCrab. 
All the XIS sensors were operated with 1/4 window option 
in order to reduce possible pile-up effects.
In this mode, the restricted detector regions corresponding
to $17^\prime{.}8 \times 4^\prime{.}5$ in the sky are read out every 2~sec.


\section{Data Reduction}
For the XIS analysis, we retrieve ``cleaned event files'', 
which are screened with standard event selections. 
We further screen the events with the following criteria:
(1) cutoff rigidity larger than 6~GV/$c$ and 
(2) elevation angle larger than $5^{\circ}$ and $20^{\circ}$ 
from the Earth rim during night and day, respectively.
We do not use XIS1 data in the following analysis 
since it suffered from telemetry saturation for 
almost the entire observation period.

The XIS events are extracted from a circular region with 
a radius of 176 pixels ($\sim 3'$) centered on the image peak. 
This extraction circle is larger than the window size 
and therefore the effective extraction region is 
the intersection of the window and this circle. 
Since the background is entirely negligible throughout 
the observations (less than 1~\% of the signal even at 9 keV), 
we do not subtract background from the data. 
%
The source was so bright during these observations 
that the XIS suffered from photon pileup at the image center 
even with the 1/4 window option.
In the present study, we exclude a circular region of the radius 
of 20 pixels at the image center from the event extraction region 
to minimize these effects. 
After excluding the central region, 
the systematic error due to the pile-up effects included 
in the flux is estimated to be less than 2~\%.

In order to take into account the complex shape 
of the event extraction region (an annulus truncated with the window size), 
we calculate the response matrices (RMF) and the effective area (ARF) 
for each XIS sensor by using {\tt xisrmfgen} and {\tt xissimarfgen} 
\citep{ishisaki06}, respectively.
The {\tt xissimarfgen} is based on Monte Carlo simulation and is designed 
to properly handle the geometrical shape of the extracted region.
Since the three FI CCDs have almost identical performance, 
we sum their data to increase statistics in the spectra. 
Accordingly, we calculate the corresponding RMF and ARF response/area files 
by adding those for XIS.

For the HXD data, ``uncleaned event files''
are screened with standard event screening criteria.
We exclude events during 
(1) SAA passages and
(2) Earth occultation, and also 
(3) those with cutoff rigidity less than 6~GV/$c$.
To estimate accurately the Non-X-ray Background~(NXB) is the key 
to realize the high sensitivity observation for HXD.
In this analysis, we use bgd\_d (for rev~1.2) 
for the NXB of HXD/PIN because the current public background model 
(bgd\_a) is found to overestimate the PIN background 
in the period during 2006 Mar 27 to May 13 \citep{fuka09}.
Note that we have to correct the effect of dead time by 
{\tt hxddtcor} for this bgd\_d as well as the source event files.
The current NXB model is shown to be accurate within $\sim$3~\% 
\citep{fuka09}.
Since the HXD/GSO detected no significant signals above 2~$\sigma$ 
assuming 5~\% systematics for the NXB estimation,
below we report results obtained from the XIS and the HXD/PIN.

\begin{figure}[t]
\begin{center}
\plotone{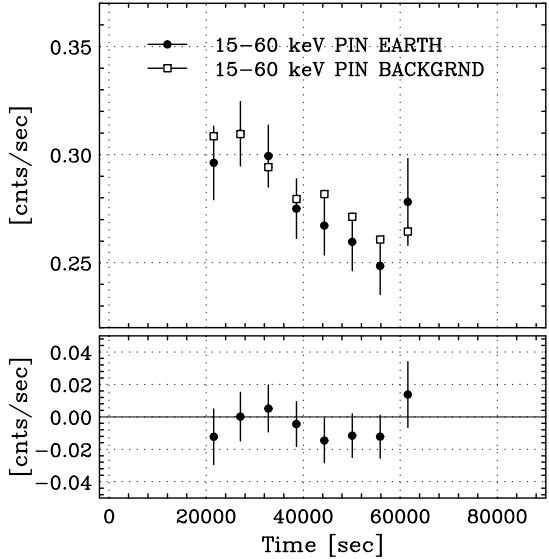}
\end{center}
\caption{Light curve of HXD/PIN during Earth occultation. 
In the upper panel, the obtained data and the background model 
in the energy band of 15--60~keV are compared.
The bottom panel shows the differences between them.}
\label{fig:earthoccultation}
\end{figure}

The accuracy of the background model is evaluated 
by comparing the count rate during the Earth occultation 
between the observation data and the NXB model,  
since the NXB model is constructed based on the Earth occultation,
which is defined by the elevation angle from the Earth rim 
less than $-5^{\rm o}$ (${\rm ELV}<-5$).
In Figure~\ref{fig:earthoccultation}, we compare the light curves 
of HXD/PIN data in 15--60~keV band actually taken 
during the Earth occultation and of the NXB model. 
Each data point is about 1~ksec accumulation. 
Figure~\ref{fig:earthoccultation} shows that the NXB model is 
well reproduced with an accuracy of less than 5~\% of the NXB even for 
1~ksec observation. 
Therefore, the uncertainty due to the misestimation of the NXB 
does not affect our results, because 
the flux is more than 30~\% of the NXB above 25~keV even 
in the lowest flux period throughout the observation, 
as we will see below.

Since the NXB model does not include contributions 
from the CXB~(Cosmic X-ray Background), 
a simulated spectrum of the CXB is added to the NXB model. 
Following \citet{gru99}, who reanalyzed HEAO-1 observations 
in the 1970s, our spectral model for the extragalactic 
background radiation is chosen as
\begin{equation}
	\frac{dN}{d\epsilon} 
	= 7.9 \;\epsilon_{{\rm keV}}^{-1.29} {\rm exp} 
	  \left( -\frac{\epsilon_{{\rm keV}}}{\epsilon_{\rm p}} \right)
	  {\rm ph\; s^{-1} keV^{-1} cm^{-2} sr^{-1}}                                  
\end{equation}
where $\epsilon_{\rm keV} = \epsilon / 1\;{\rm keV}$ 
and $\epsilon_{\rm p} = 41.13$.
We estimate the expected CXB signal in the HXD/PIN spectrum 
using the latest response matrix for spatially uniform emission, 
{\tt ae\_hxd\_pinflate1\_20070914.rsp}. 
The contribution from the CXB flux is estimated to be 5~\% 
of the NXB, which is comparable to the current systematic uncertainties 
of the NXB model itself.

\section{Analysis}
\subsection{Light Curves}

\begin{figure}[t]
\begin{center}
\plotone{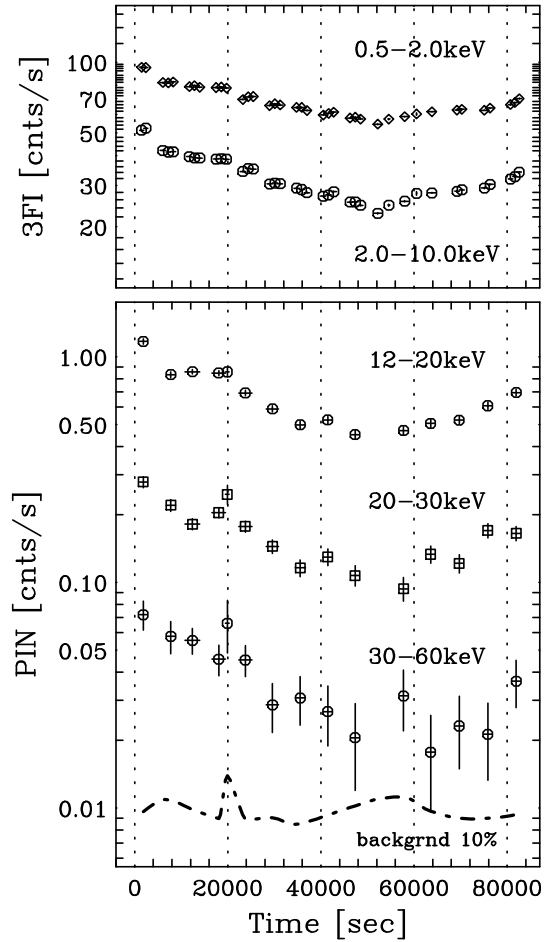}
\end{center}
\caption{The sum of data from three XIS sensors and HXD/PIN light curves 
in the energy ranges of 0.5--2.0~keV, 2.0--10.0~keV, 12--20~keV, 
20--30~keV and 30--60~keV. The dot-dash line corresponds 
the 10\% of NXB for 30--60~keV band.}
\label{fig:lightcurve}
\end{figure}

The time history of {\it Suzaku} observation obtained with XIS 
and HXD is shown in Figure~\ref{fig:lightcurve} for five energy bands.
The flux in each energy band decreases by factor 2--4 
during the first half of the observation and then turns 
into an increase starting at around the middle of the observation. 
Thanks to low background characteristics of HXD, 
the variability with time scale of as short as 20~ksec is 
clearly detected even at the highest band, 30--60~keV.
The dot-dash line drawn in the bottom of the figure represents 
the 10~\% of the estimated NXB 
for the highest energy band, whereas the contribution from CXB 
in the same energy band is about 0.002 counts/sec.
This also supports the significant detection of the signal 
from Mrk~421 up to 60~keV in such short time intervals.

The change in flux becomes larger in higher energy bands, 
which is consistent with previous studies of Mrk~421 
\citep[e.g.,][]{tad96, tad00}.
The count rate decreases from 100~count s$^{-1}$ down to 
70~count s$^{-1}$  in the lowest energy band (0.5--2.0~keV in the XIS data), 
while it drops from 0.07 counts/sec down to 0.02 counts/sec 
in the highest energy band (30--60~keV in the HXD data), 
that is, the higher energy band is more than twice 
as variable as the lower energy band,
even considering the systematic and statistical errors.

\subsection{{\it Suzaku} Spectral Analysis} \label{subsec:widebandana}
%
\begin{figure}[b]
\begin{center}
\plotone{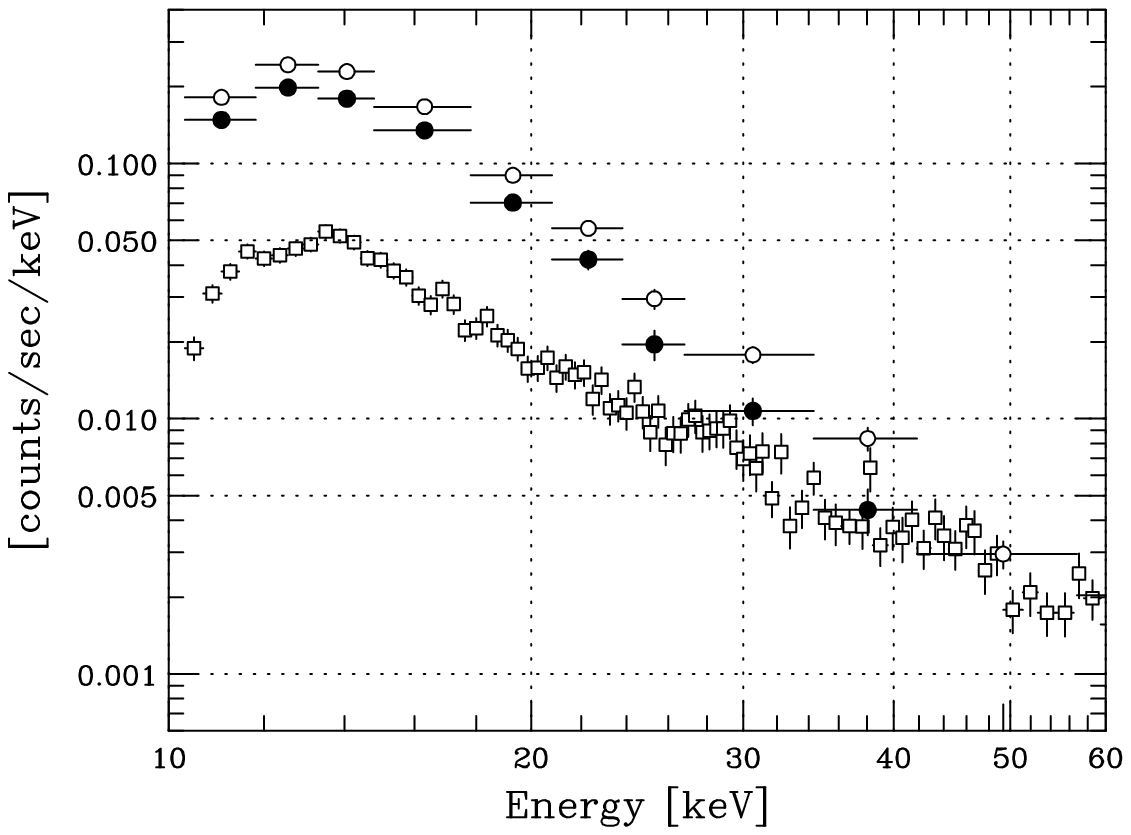}
\plotone{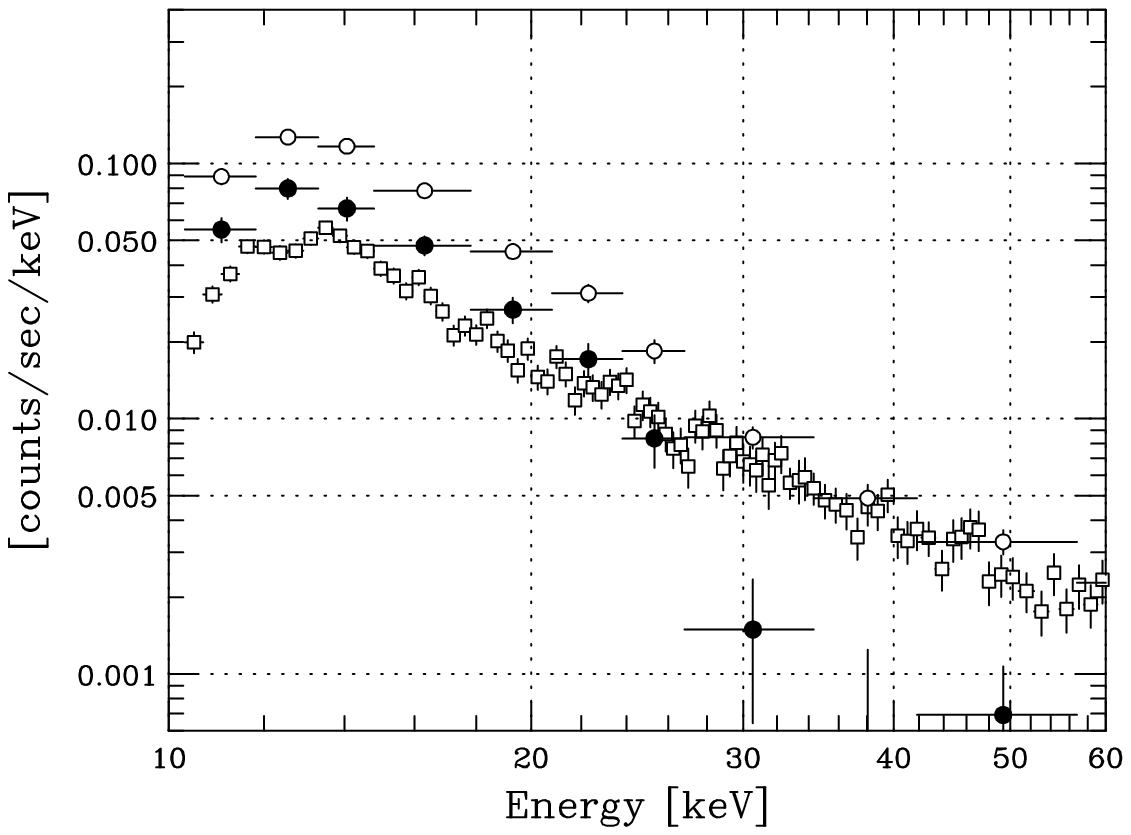}
\end{center}
\caption{The raw spectra of HXD/PIN at the brightest period (upper panel) 
and at the lowest period (lower panel). 
Only 750 sec accumulation time is used for the brightest period 
and 1.7 ksec for the lowest period. 
The open circles, open squares and filled circles show the raw data, 
background (NXB + CXB), and background-subtracted data, respectively.}
\label{fig:rawspectra}
\end{figure}

In order to study the time evolution exhibited 
in the light curves more quantitatively, 
we divide the XIS and HXD data into 14 time intervals 
and perform spectral fitting to XIS+HXD spectra 
derived from each time interval.
Raw HXD/PIN spectra during the highest 
and the lowest flux periods are shown in Figure~\ref{fig:rawspectra}.
During the observation, the flux below 20~keV is 
always higher than that of the NXB.
At the lowest flux period (lower panel of Figure~\ref{fig:rawspectra}), 
flux level is comparable to the NXB around 25~keV 
and about 30~\% of the NXB level above 25~keV.
Since the systematic error of the NXB is about 5~\% \citep{fuka09}, 
which is six times less than 
the flux level measured at the lowest flux period,
the results of present analysis are not affected 
by the systematic uncertainties of the NXB.
In the following spectral analysis, the column density 
is fixed to the Galactic value: $N_H = 0.0161 
\times 10^{22} ~{\rm cm}^{-2}$ \citep{lockman95} 
and the cross normalization between XIS and HXD/PIN 
is also fixed to 1.13 \citep{kokubun06}.

\begin{figure}[t]
\begin{center}
\plotone{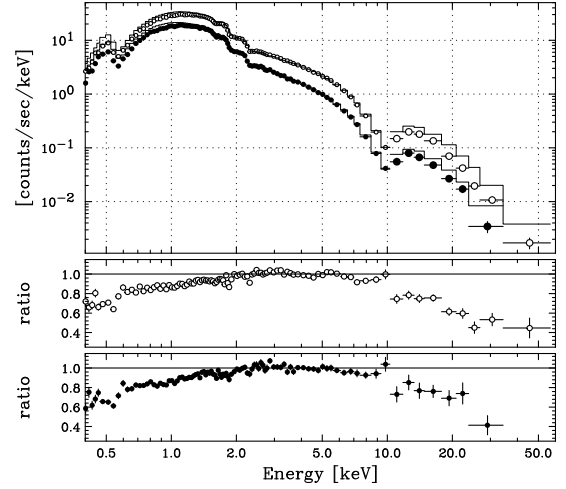}
\end{center}
\caption{XIS$+$PIN spectra in the energy range 0.4--60 keV 
are shown with the best-fit models obtained by fitting 
the XIS spectrum (2--5~keV) alone with power law model.
The data points below/above 10~keV correspond to XIS and PIN data respectively.
The lower panels show the ratio between the data points 
and the model values for the highest and the lowest periods.}
\label{fig:wide}
\end{figure}

Wide band spectrum obtained by {\it Suzaku} enables us 
to measure time-resolved behavior of the continuum spectrum, 
including even small deviations from a simple power law model.
Firstly, in order to see how the spectra deviate 
from power law function, we restrict the fitting range 
to 2--5~keV.
The results of fitting are shown in Figure~\ref{fig:wide}, 
and as is clearly seen in the ratio, the spectra gradually 
curve downwards for both cases, meaning the power law index increases 
with increasing energy.
Such a departure from a simple power law model appears 
to be a general feature of X-ray spectra of BL-Lac type blazars 
with the peak of the synchrotron component in the UV to 
soft X-ray band \citep[see, e.g.,][]{perlman05}.
It is apparent that the spectrum changes with time, 
even in this restricted energy range: 
the spectral indices of power law model (2--5~keV) significantly 
change from 2.05$\pm$0.02 to 2.27$\pm$0.03,
with the flux-index correlation in the sense 
where the soft X-ray spectra soften as the flux decrease.
This is consistent with the results published so far 
\citep[e.g.,][]{tad96}.
Clearly, sensitive {\it Suzaku} observations indicate spectral 
variability on short time scales (a few hour), and provide 
an opportunity to derive important constraints in the context of 
more physical ``synchrotron model,'' where the data are directly 
fitted to a particle spectrum radiating via the synchrotron process.
This treatment is covered in the following sections.

\subsection{Interpreting the X-ray Spectrum 
in the Context of Physical Synchrotron Model} \label{subsec:synchrotron}
We can obtain constraints on the physical parameters 
of the synchrotron emission, such as $B$ and $\gamma_{\rm max}$, 
via fitting the spectra with the recent numerical model \citep{ttanaka08}.
Hereafter we simply call this model ``synchrotron model.''
In the model, we assume an electron spectrum has 
a form of power law with exponential-type cutoff 
at the maximum energy $E_{\rm max}$ (= $\gamma_{\rm max} m_ec^2$), namely,
\begin{equation}
	\frac{dN_e}{dE}	 = N_{e0} E^{-s} {\rm exp}
	\left[ - \left(\frac{E}{E_{\rm max}}\right)^\beta  \right]. 
	\label{eq:dist}
\end{equation}
Here, $s$ is the energy index of electron distribution, while 
$\beta$ and $N_{e0}$ (electrons cm${}^{-3}$ eV${}^{-1}$) determine 
the shape of high energy end of the population and the normalization, 
respectively.
As described in \citet{kataoka99}, we assume the spherical 
and uniform emitting region. 
The total power per unit frequency $P(\epsilon)$ can be written as
\begin{equation}
	P_{\rm tot}(\epsilon) \propto \int 
	F\left( \frac{\epsilon}{\epsilon_c} \right) \frac{d N_e}{dE} dE.
\end{equation}
Here $\epsilon$ and $E$ parameters indicate the photon 
and electron energy, respectively, and the function $F(x)$ is 
defined as
\begin{equation}
	F(x) \equiv x\int_x^\infty K_{5/3}(\xi)d\xi,
\end{equation}
where $K_{5/3}(\xi)$ is the modified Bessel function of 5/3 order.
When pitch angles are isotropic, the characteristic photon energy, 
$\epsilon_c$, is given as
\begin{equation}
	\epsilon_c	= 5.43~\left( \frac{B}{0.10~{\rm G}} \right)
	\left( \frac{E}{1~{\rm TeV}} \right)^2~{\rm keV}.
\end{equation}
This model has three free parameters; $s$, $\Pi \equiv E_{\rm max}{B}^{1/2}$~
[GeV G${}^{1/2}$] and the normalization of a spectrum.
We note here that this synchrotron model is in fact based 
on the distribution of energies of radiating particles 
that is essentially a power law with an exponential 
(or hyper-exponential) cutoff.
Such a distribution is similar to that which has been calculated 
to stochastic acceleration via the Fermi II process 
- such as, for instance, via scattering 
in magnetic turbulent regions in plasma 
\citep[see, e.g.,][]{stawarz08}.

In the case of emission from a jet, we need to consider 
the effect of relativistic beaming,
since the fitted value of $\Pi$~($\Pi_{\rm obs}$) is not given 
in the jet's frame ($\Pi_{\rm jet}$),
but instead $\Pi_{\rm obs} = \Pi_{\rm jet} \times \delta^{1/2}$.
We adopt the Doppler (beaming) factor $\delta = 10$: 
this value was directly measured by VLBI observations 
\citep[e.g.,][]{verm94} 
which is consistent with the result obtained by simultaneous 
multi-band analysis from radio to $\gamma$-ray \citep{kubo98}.  
We note that more recent work by \citet{lichti08}
which analyzes the big outburst followed by some weeks with respect to 
our {\it Suzaku} observation and suggests $\delta = 15$ by modeling the SED. 
However, these differences are not the scope of our paper and  
will not affect our discussion. 

\begin{figure}[b]
\begin{center}
\plotone{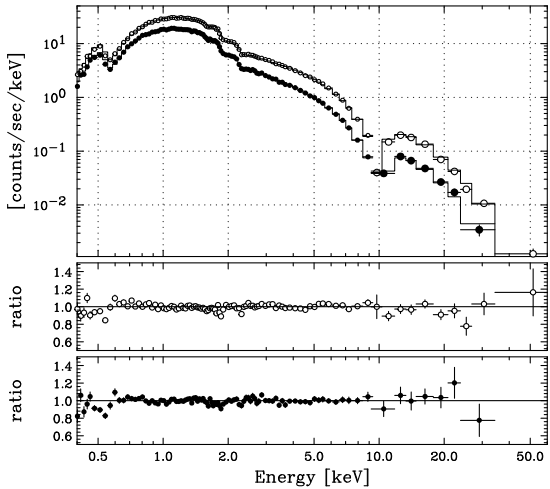}
\end{center}
\caption{The same as Figure~\ref{fig:wide} 
but fitting with simple synchrotron model 
in entire energy band 0.4--60~keV.}
\label{fig:wide_sync}
\end{figure}

\begin{figure}[htb]
\begin{center}
\plotone{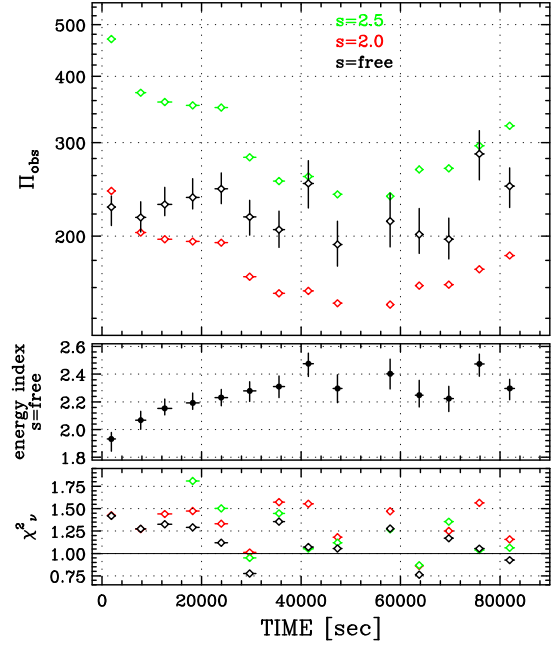}
\end{center}
\caption{The results of fitting time-resolved spectra 
with simple synchrotron model. 
The top panel represents the evolution of $\Pi_{\rm obs}$ values 
in units of GeV G${}^{1/2}$ 
obtained by fitting each spectrum in the case of the energy index 
of electron spectrum fixed to $s=2.0$ (red), $s=2.5$ (green) 
and $s={\rm free}$ (black).
The error bars include only statistical $1\sigma$ errors and those
for green and red diamonds are too small to be seen.
The middle panel shows the history of the energy index 
of electron spectrum in the case of $s={\rm free}$.
In the bottom panel, we show the $\chi_\nu^2$ values corresponding to 
59 and 60 d.o.f.}
\label{fig:fitting_result}
\end{figure}

\begin{figure*}[htb]
\begin{center}
\plotone{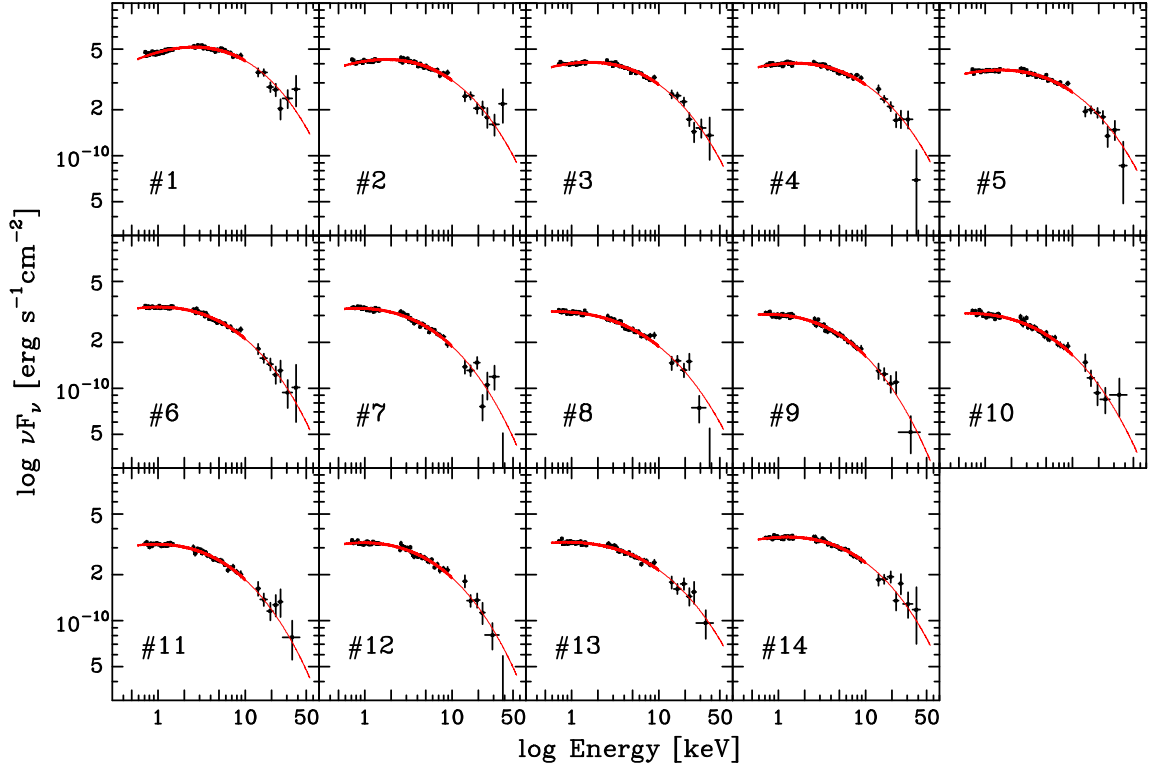}
\caption{SED for each good time interval fitted by ``synchrotron`` model.}
\label{fig:SED_all}
\end{center}
\end{figure*}

When we perform the spectral fitting for each data set 
with this model with $\beta$ fixed to unity, 
almost all the individual spectra are 
well fitted by this synchrotron model. 
In the fitting, we exclude the bandpass 1.5--2.5~keV and 8.0--10.0~keV 
because there remains some calibration uncertainties of the XIS detectors.
Moreover, we fix the energy index of electron distribution ($s$) to be 2.0:
then, the reduced chi-square ($\chi_\nu^2$) is ranging from 0.86 to 1.57 
with 57 degrees of freedom (dof).
In some time intervals, the $\chi_\nu^2$ values are 
substantially larger than in others and the model will be rejected. 
However, relatively large values of $\chi_\nu^2$ 
are partially caused by the insufficient calibration of the XIS/XRT 
rather than the inappropriate modeling of the spectra.
Figure~\ref{fig:wide_sync} shows the typical examples 
of the results for the highest- and the lowest-flux periods.
For those, we obtain $\Pi_{\rm obs}$ = $243\pm 2$ ($\chi_\nu^2=1.42$ for 57 dof) 
and $149\pm 1$ ($\chi_\nu^2=1.18$ for 57 dof)~[GeV G${}^{1/2}$] 
for the highest and the lowest period, respectively.

We summarize the fitting results of each spectrum 
by the ``synchrotron'' model in Figure~\ref{fig:fitting_result}.
The top panel of Figure~\ref{fig:fitting_result} shows 
the evolution of $\Pi_{\rm obs}$ as a function of time, 
where the data points in red color correspond to 
the case of $s=2.0$.
With $s=2.0$, the $\Pi_{\rm obs} = E_{\rm max}B^{1/2}\delta^{1/2}$ varies 
largely as the flux changes. 
In order to study how the energy index of electrons affects 
the results, we also fit those spectra with $s=2.5$, 
which is shown by the green markers 
in the Figure~\ref{fig:fitting_result}.
The $\Pi_{\rm obs}$ values in the case of $s=2.5$ are 
significantly different (higher) from those in the case of $s=2.0$.
In both cases, $\Pi_{\rm obs}$ parameters change by a factor of 2.
On the other hand, when we treat $s$ as a free parameter, 
$\Pi_{\rm obs}$ does not change significantly, 
while $s$ increases from 1.95 to 2.55 in the decreasing phase.
Statistically, $\chi_\nu^2$ for the case of $s$ as free is 
the lowest in any time intervals compared with in the case of fixed $s$ 
by only an additional degree of freedom, 
except for the first and the second time intervals.
The interpretation of those results is discussed in \S~\ref{disc:pi_max}.

The spectra are fitted fairly well with the model based on 
synchrotron emission.
In Figure~\ref{fig:SED_all}, we plot the spectra in $\nu$--$\nu F_\nu$ 
space for all the data sets.
All the spectra have convex shape.
The shape of these spectra change significantly, 
and that is quite pronounced in the first interval (\#1).
In the first half of the observation, the flux decreases 
as the synchrotron peak frequency gradually shifts 
to lower band, that is, to $\sim 5$~keV 
at the highest period (\# 1) and below 0.4~keV 
at the lowest period (\# 9).
On the other hand, in the second half, 
the synchrotron peak frequency does not seem to change apparently,
as the flux increases.

Since X-ray spectra of blazars are often fitted to a spectral 
model involving a broken power law, for comparison, we 
attempted such a model.  In all cases, the broken power law 
(involving three free parameters plus normalization, namely index before the 
break, after the break, and the break energy) returns a worse fit than 
the synchrotron model (where we used $s$ fixed at 2):  for instance, 
for period \#~1, $\chi_\nu^2$ is 1.42 for 57 dof for 
the ``synchrotron'' model while it is 2.28 for the broken power law;  
for period \#2, the values are respectively 1.27 vs. 1.40:  
the broken power law model appears to have a break that 
is too sharp to fit the data.  
As it will become important in the discussion below, 
most relevant is the comparison of the two models for the period \#~9:  
here, the values are respectively 1.18 vs. 1.90. 

\subsection{Spectral Analysis of the Steady and the Variable Components} 
\label{subsec:variablecomp}
%
\begin{figure}[b]
\begin{center}
\plotone{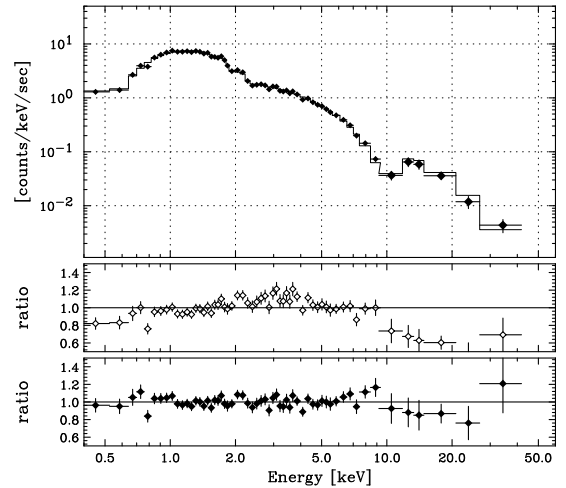}
\end{center}
\caption{Power law and broken power law fit for a subtracted spectrum. 
This is a case of second time interval (\# 2) out of 14. 
The middle and bottom panels represent the ratio between data and model 
for simple power law and for broken power law model.}
\label{fig:spec_powerlaw_sample}
\end{figure}
\begin{figure}[b]
\begin{center}
\plotone{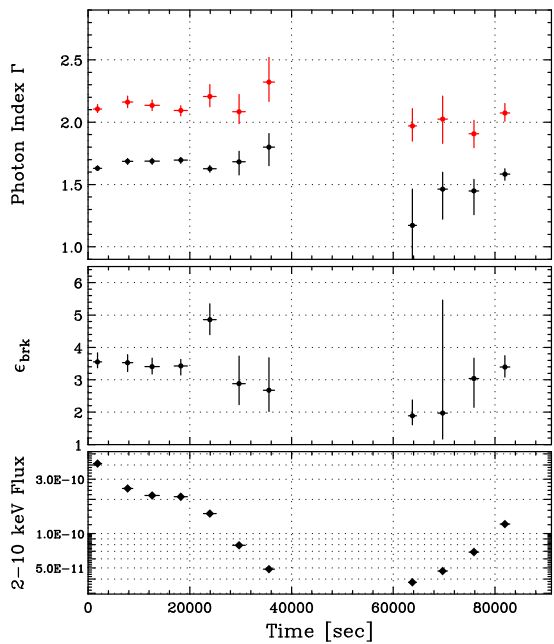}
\end{center}
\caption{Evolution of ``variable'' spectra. Top and middle panels 
show the history of the photon indices for lower (black; $\Gamma_{\rm lo}$) 
and higher (red; $\Gamma_{\rm hi}$) 
energy band and the break energy ($\epsilon_{\rm brk}$) in units of keV. 
We consider the ``steady'' component to correspond to the lowest-flux state (\#~9).
The vicinity of \#~9, that is, \#~8 and \#~10 are excluded because of relatively 
poor statistics.
Bottom panel shows that of the 2--10~keV flux in units of 
erg s${}^{-1}$cm${}^{-2}$.}
\label{fig:lc_bknpower}
\end{figure}
\begin{deluxetable*}{cccccc}
\tabletypesize{\scriptsize}
\tablecaption{
Fit results of the ``variable'' components with a broken power law function 
\tablenotemark{a} 
\label{table:bknpower}
}
\tablehead{
\colhead{\#}	&	\colhead{$\Gamma_{\rm lo}$\tablenotemark{b}}	&	
\colhead{$\epsilon_{\rm brk}$\tablenotemark{c}}&	
\colhead{$\Gamma_{\rm hi}$\tablenotemark{b}}	&
\colhead{2--10~keV flux\tablenotemark{d}}		&	\colhead{$\chi_\nu^2$~(dof)}
}
\startdata
1 & $1.63^{+0.02}_{-0.02}$  &  $3.6^{+0.3}_{-0.2}$  
&   $2.11^{+0.04}_{-0.03}$  &  $4.12\times 10^{-10}$  & $1.25$ (54)\\
2 & $1.69^{+0.02}_{-0.02}$  &  $3.5^{+0.2}_{-0.3}$  
&   $2.16^{+0.05}_{-0.04}$  &  $2.49\times 10^{-10}$  & $1.08$ (54)\\
3 & $1.69^{+0.02}_{-0.03}$  &  $3.4^{+0.3}_{-0.2}$  
&   $2.13^{+0.04}_{-0.04}$  &  $2.16\times 10^{-10}$  & $0.87$ (54)\\
4 & $1.70^{+0.02}_{-0.03}$  &  $3.4^{+0.2}_{-0.3}$  
&   $2.09^{+0.04}_{-0.04}$  &  $2.10\times 10^{-10}$  & $1.23$ (54)\\
5 & $1.63^{+0.03}_{-0.03}$  &  $4.9^{+0.5}_{-0.5}$  
&   $2.21^{+0.10}_{-0.08}$  &  $1.50\times 10^{-10}$  & $0.97$ (27)\\
6 & $1.68^{+0.09}_{-0.10}$  &  $2.9^{+0.8}_{-0.7}$  
&   $2.08^{+0.14}_{-0.10}$  &  $7.90\times 10^{-11}$  & $0.81$ (27)\\
7 & $1.80^{+0.11}_{-0.15}$  &  $2.7^{+1.0}_{-0.7}$  
&   $2.32^{+0.20}_{-0.16}$  &  $4.88\times 10^{-11}$  & $1.17$ (27)\\

\\

11 & $1.17^{+0.29}_{-0.35}$  &  $1.9^{+0.5}_{-0.3}$  
&    $1.97^{+0.14}_{-0.12}$  &  $3.75\times 10^{-11}$  & $0.72$ (27)\\
12 & $1.46^{+0.14}_{-0.24}$  &  $2.0^{+3.5}_{-0.8}$  
&    $2.02^{+0.18}_{-0.20}$  &  $4.70\times 10^{-11}$  & $1.14$ (27)\\
13 & $1.45^{+0.09}_{-0.19}$  &  $3.0^{+0.6}_{-0.9}$  
&    $1.91^{+0.11}_{-0.11}$  &  $6.91\times 10^{-11}$  & $0.75$ (54)\\
14 & $1.58^{+0.04}_{-0.05}$  &  $3.4^{+0.3}_{-0.3}$  
&    $2.07^{+0.08}_{-0.07}$  &  $1.21\times 10^{-10}$  & $0.80$ (54)
\enddata
\tablenotetext{a}{The evolution of these parameters is shown in Figure~~\ref{fig:lc_bknpower}.} 
\tablenotetext{b}{The $\Gamma_{\rm lo}$ and $\Gamma_{\rm hi}$ indicate 
the photon indices below and above the break energy $\epsilon_{\rm brk}$.}
\tablenotetext{c}{The break energy $\epsilon_{\rm brk}$ is in units of keV.}
\tablenotetext{d}{The 2--10~keV flux is in units of erg s${}^{-1}$ cm${}^{-2}$.}
\end{deluxetable*}

\begin{figure*}[htb]
\begin{center}
\plotone{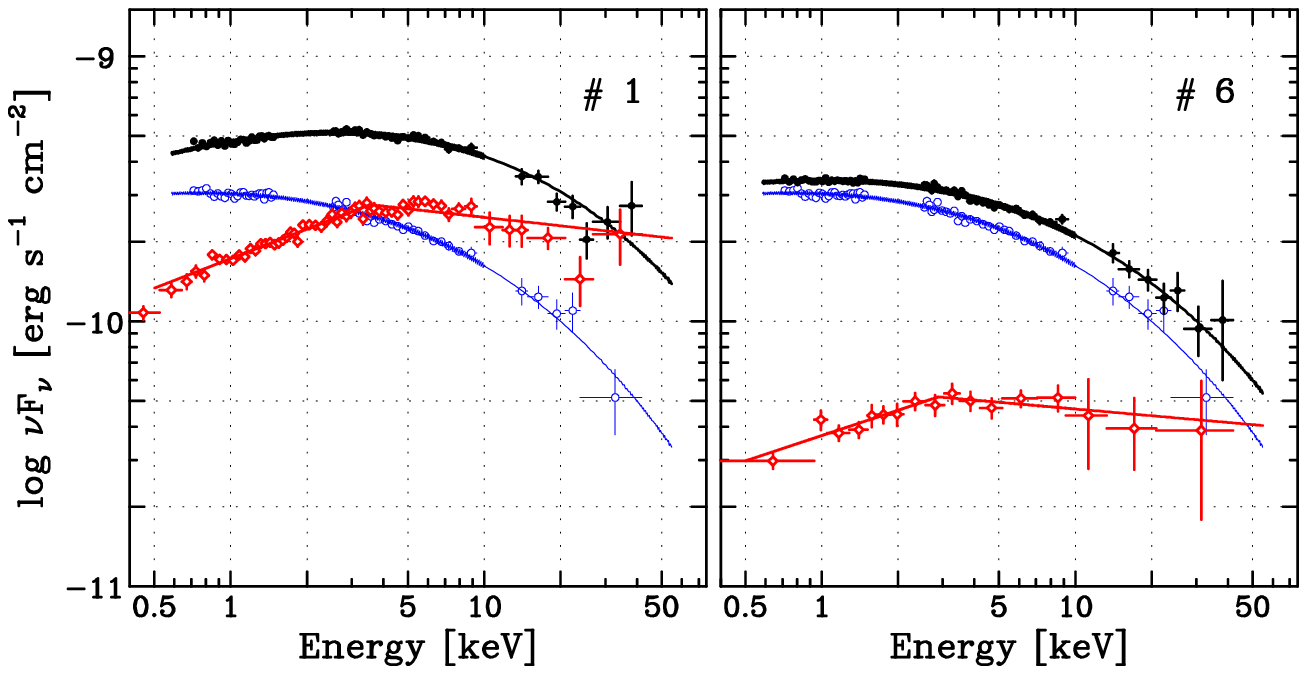}
\end{center}
\caption{Simultaneous plots of ``variable'' and ``steady'' 
components for the highest flux period (epoch \#~1, left panel) 
and the end of decay period (epoch \#~6, right panel). 
The red and black data set are for the ``variable'' components 
and the total (=``steady''+``variable'') spectra.
The blue data set corresponds to the lowest-flux period and is 
common to two plots.}
\label{fig:SED_variable_steady}
\end{figure*}

One possible approach towards the exploration of 
the spectral evolution during the observation is 
to separate the X-ray data into a ``steady'' and a ``variable'' components. 
In reality, ``steady'' means slowly-variable, 
variable on time scales much longer than typical X-ray observations 
lasting for one or several days, since 
historically the source flux does drop 
well below the level of our ``steady'' level.
We associate the ``variable'' component with a newly injected 
electron distribution which drives the flare activity.
We consider the ``steady'' component to correspond to 
the lowest state spectrum (\#~9), 
which is illustrated in Figure~\ref{fig:rawspectra} and well described by 
the synchrotron model in Figure~\ref{fig:wide_sync},
and the ``variable'' component 
to be the difference of individual spectra obtained 
at other epochs {\it minus} that of the lowest state spectrum.

In order to unveil the characteristics of ``variable'' component, 
we fit each of the subtracted spectra with simple power law model.
The photon index of the subtracted spectrum 
is found to be harder, $\Gamma \sim 1.8$, 
but it also shows a large discrepancy between data and model 
and therefore is not well fitted by this model.
Next we attempt a broken power law model 
and find that this model is acceptable for all the time intervals.
An example of the fitting is shown 
in Figure~\ref{fig:spec_powerlaw_sample} comparing 
with that of simple power law model, corresponding to the epoch \#~2, 
and $\chi_\nu^2$ is significantly improved 
from 3.13 (56~dof) to 1.08 (54~dof).
Subsequently, we fit all of the intervals 
with such a broken power law spectrum :
Figure~\ref{fig:lc_bknpower} shows the time history of parameters, 
that is, the photon index of low energy band ($\Gamma_{\rm lo}$) 
and of high energy band ($\Gamma_{\rm hi}$), and the break energy 
($\epsilon_{\rm brk}$), which are summarized in Table~\ref{table:bknpower}.
The vicinity of the epoch \#~9 are excluded because the relatively large
statistical errors of energy bins make it difficult to extract meaningful values.
The 2--10~keV flux of the ``variable'' component 
decreases from $4.1\times 10^{-10}$ to 
$4.9\times 10^{-11}$ erg s${}^{-1}$cm${}^{-2}$ (and of course
to zero for interval \#~9, by construction) 
with the e-folding decaying time of $1.7\times 10^4$ sec, 
that is, less than 5 hours.

These plots lead to two important results : 
(1) the spectral shape of ``variable'' component does not vary
across entire observation and only the normalization changes
and 
(2) the photon indices $\Gamma_{\rm lo}$ and $\Gamma_{\rm hi}$ are 
about 1.6 and 2.1 respectively, that is, 
the differences between them are equal to about 0.5,
with the break energy in photon space remaining relatively constant, 
at $\epsilon_{\rm brk} \simeq 3$~keV.
All this has important implications which we discuss 
in \S~\ref{disc:variable}.

It is interesting to compare the lowest spectrum 
- that corresponding to the \# 9 epoch - to the remaining spectra.
This is essentially a comparison of the ``variable'' 
and ``steady'' components.
In Figure~\ref{fig:SED_variable_steady}, we plot both spectra 
at the same graph in the $\nu$--$\nu F_\nu$ space;  
there, we illustrate the highest flux period (\# 1 minus \# 9) 
and the end of decaying period (\# 6 minus \# 9) 
for the ``variable'' components.
The spectral shapes of the ``steady'' and ``variable'' components are 
quite different, specifically the spectral indices of 
``variable'' components, both lower and higher energy band, 
are harder than these of ``steady'' component.
Perhaps the most striking is the fact that 
the spectral shapes of ``variable'' components 
do not vary significantly, and only the normalization changes.

\section{Discussion}
In the previous section, we performed detailed spectral analysis 
of TeV blazar Mrk~421 obtained by {\it Suzaku} in the period of 
the large flare in 2006 April 28--29 
and found that there are two different interpretations 
for their variability.
Below, we consider the physical scenarios that might correspond 
to the two phenomenological models above.

\subsection{Spectral Variability due to the Change of $\Pi_{\rm obs}$}
\label{disc:pi_max}

In \S~\ref{subsec:synchrotron}, we successfully fitted the evolution 
of the X-ray spectrum by the synchrotron model.
In the case where the energy index of electron population is allowed to be 
free, $\Pi_{\rm obs}$ parameter does not change significantly 
throughout the observation and the change of electron energy index $s$ 
appears to be the factor responsible for the spectral change.
The electron energy index itself is unlikely to vary significantly, 
so if the spectral evolution is due to the change 
of single electron population, the observed variability 
must be due to the change of $\Pi_{\rm obs}$.
 
Using equation (2.31) in \citet{inoue96}, $\Pi_{\rm obs}$ 
can be written as
\begin{equation}
	\Pi_{\rm obs} = B^{1/2} \delta^{1/2} \gamma_{\rm max} m_e c^2 
  = v_sBm_ec \left[ \frac{9e\delta}{80(u_B+u_{\rm sync}) 
	\sigma_{\rm T} \xi} \right]^{1/2}
	\label{eq:pi_obs}
\end{equation}
where $v_s$, $u_B$, $u_{\rm sync}$, $\sigma_{\rm T}$ and 
$\xi$ are the shock velocity in the observer's frame, 
the energy density of magnetic field and synchrotron seed photons, 
Thomson cross section, and the so-called gyro-factor 
(the ratio of the mean free path for scattering with magnetic 
disturbances compared to the Larmor radius), respectively. 
This equation assumes that the maximum energy of electron distribution is 
determined by equating the radiative cooling time 
to the accelerating time. 
From this equation, the variability of $\Pi_{\rm obs}$ can be 
explained by changes of one or more parameters such as $v_s$, 
$B$ (thus also $u_B$), $u_{\rm sync}$, $\xi$, or $\delta$. 
The degeneracy between the parameters can be broken by comparing 
the temporal and spectral behavior of the X-ray time series against the 
TeV flux history.  
For instance, if the magnetic field $B$ is mainly responsible for the 
changes of $\Pi_{\rm obs}$, it would require for the ratio 
of the synchrotron luminosity to the inverse Compton luminosity,  
$L_{\rm sync}/L_{\rm IC}$, significantly change as a function 
of time. 
However, such multi-wavelength studies require simultaneous  
X-ray and TeV data, which are not currently available for this
{\it Suzaku} observation.  

\subsection{Spectral Variability due to Superposition of Steady 
and Variable Components}  \label{disc:variable}
As an alternative and preferred scenario to explain the variable behavior 
of Mrk~421, we introduce two separate electron populations, 
one is ``steady'' and the other is ``variable'' corresponding 
to the steady and variable componsnts of the X-ray spectra.  
There, a fresh electron distribution, distinct than  
the pre-existing one, is newly injected into 
the same (or a different) region of the jet and is responsible 
for the observed variable portion of source flux.  

One of the important results in \S~\ref{subsec:variablecomp} is 
that the photon indices $\Gamma_{\rm lo}$ and $\Gamma_{\rm hi}$ are 
measured to be $\sim 1.6$ and $\sim 2.1$ for ``variable'' components.
Surprisingly, the result suggests that electrons form the power law 
distribution of energy index $\sim 2$, 
which is predicted from the standard shock acceleration theory 
for both non-relativistic and relativistic case 
\citep{bland78, kirk00} and then suffer from synchrotron cooling. 
As for the variability, we have shown that 
the spectral shape of ``variable'' components does not vary 
throughout the observation and the break energy appears to stay 
constant around $\epsilon_{\rm brk}\sim 3$~keV.
While it is possible to adjust other parameters simultaneously 
to keep $\epsilon_{\rm brk}$ constant, this would require simultaneous 
changes of several parameters, amounting to ``fine-tuning`` of parameters.
The simplest interpretation of this observational result is that 
the variability of the X-ray flux is only due to the change of 
the injection rate of electrons, $N_{e0}$ in Eq.~(\ref{eq:dist}).
Since the break energy in the synchrotron spectrum ($\epsilon_{\rm brk}$) 
is constant, $B$, $\delta$ and $\gamma_{\rm brk}$ are also expected 
to stay constant.  With this, the synchrotron luminosity is proportional 
only to the energy density of the particles:
\begin{equation}
	L_{\rm sync}\propto u_eu_B\delta^4 
				\propto u_e\delta^2\gamma_{\rm brk}^{-4} 
				\propto N_{e0}. \label{eq:Lsync}
\end{equation}

We thus decompose the X-ray spectrum of Mrk~421 
into a ``steady'' component described as due to 
an exponentially cutoff power law distribution of electrons, 
and the ``variable'' component, well-described 
as a broken power law photon spectrum.  
Here we can make a suggestion of physical picture for 
these two components: 
the ``variable'' component is due to electrons efficiently 
accelerated in relatively small regions, such as 
localized shocks, via Fermi I process.  
In addition to that, a relatively ``steady'' component 
contributes as well: 
one possibility would be that the observed spectrum 
of this component is produced by shocks located at 
larger distance along the jet from the 
black hole.
Such picture, invoking an internal shock scenario, 
was suggested by \citet{tanihata03}.  There, 
the resolved, rapid flares are due to collisions of 
pairs of shells 
at a characteristic distance $D_{\rm ch}$ 
from central engine, while those colliding at larger distance than 
$D_{\rm ch}$ make up the underlying, slowly variable component.  
Alternatively, the ``steady'' component 
might be associated with another 
process: the electrons escaping from the shock region 
into more extended volume are re-accelerated 
via Fermi II process on turbulence sites, as would be in the 
scenario described by several authors 
\citep{virtanen05,katar06,stawarz08}.
This process operates in a larger volume, 
and thus the longer variability time scale.
This scenario is particularly 
attractive as it is strongly supported by the spectral shapes 
and relative variability time scales of the two putative components.
Specifically, the photon spectrum calculated for 
the diffuse acceleration process by \citet{stawarz08} 
is similar to that 
derived via the synchrotron model, described and applied by us above.  

With regard to the ``variable'' component, we infer 
an important implication of the fact that 
the observed break energy seems not to change throughout the observation.
There, the break seen in the photon spectrum might be 
due to a competition between acceleration and cooling 
of relativistic electrons and expressed as
\begin{equation}
	\gamma_{\rm brk} 
	= \frac{3m_ec^2}{4(u_{\rm B}+u_{\rm sync}) \sigma_{\rm T} R}
	\label{eq:gamma_brk}
\end{equation}
where $R$ is the size of emitting region \citep{inoue96}.
To be exact, $u_{\rm sync}$ in Eq.(\ref{eq:gamma_brk}) is
actually an increasing function of $N_{e0}$ via Eq.~(\ref{eq:Lsync}) 
and this implies $\epsilon_{\rm brk}$ should increase
as the flux of ``variable'' component decreases.
However, the fact that $\epsilon_{\rm brk}$ does not seem to change
throughout the decaying phase of ``variable'' component
suggests that the energy density of magnetic field is
larger than that of soft seed photons ($u_B \gg u_{\rm sync}$)
for ``variable'' components.
This in turn suggests the synchrotron loss is dominant over the
inverse Compton loss ($u_B/u_{\rm sync} 
= L_{\rm sync,var}/L_{\rm IC,var} \gg 1$).

\section{Summary}
We performed a detailed analysis of bright TeV blazar Mrk~421 data 
collected with {\it Suzaku} during the large flare in 2006 April 28--29. 
During the observation, the flux varied from $3.7\times10^{-10}$ to 
$7.8\times 10^{-10}$ erg s${}^{-1}$cm${}^{-2}$.
Thanks to the high sensitivity of {\it Suzaku}, 
we obtain the time-resolved spectrum of wide energy band 0.4--60~keV 
in intervals as short as $\sim$ 1~ksec.

Our detailed, time-resolved spectral analysis
leads to two different scenarios describing the variability of Mrk~421.
One is the conventional, one-component picture: 
in order to interpret the spectrum, 
we introduce the synchrotron model for the fitting function 
and find that the variability is due to the change of $\Pi_{\rm obs}$, 
the products of $B^{1/2}$, $\delta^{1/2}$ and $\gamma_{\rm max}m_ec^2$. 
The variability of $\Pi_{\rm obs}$ can be
explained by changes of one or more of them.
However, since we cannot break the degeneracy only this observation, 
simultaneous X-ray and TeV $\gamma$-ray observation are required.
The other scenario invokes also a second, separate electron distribution 
that is responsible for the variability and which has energy index $s\sim 2$;  
this distribution is distinct than the one responsible for the steady component.
Here, the rapidly variable component might be due to localized 
shock (Fermi I) acceleration, while the more steady component might be 
due to the superposition of shocks located at larger distance 
along the jet or due to a larger-scale, stochastic (Fermi II) acceleration 
on turbulence sites in the shocked plasma.  

It is not possible to distinguish which scenario correctly 
describes the emission mechanism solely via this observation.
However, the multi-component scenario provides a meaningful hint 
to disentangle the puzzle of the emission mechanism, 
such as the ``orphan'' flares and the very short time scale 
($\sim$ minutes) of variability observed in TeV $\gamma$-ray band.
In order to investigate the emission mechanism of TeV blazars further, 
simultaneous broad-band, high-sensitivity observations are needed, 
especially in X-ray and GeV/TeV $\gamma$-ray band 
facilitating a comparison of the ``variable'' components 
over the entire broad-band blazar spectra.  
Those show a promise towards a progress for understanding 
the acceleration mechanism in the relativistic jet.
Such observations will be possible using {\it Suzaku}, 
{\it Fermi} GeV $\gamma$-ray telescope 
and advanced Cherenkov $\gamma$-ray telescopes.

\acknowledgments
We acknowledge financial support from NASA grant NNX08AZ89G, 
and by the Department of Energy contract to SLAC no. DE-AE3-76SF00515.
We thank the H.E.S.S. and MAGIC collaboration 
for the coordination of the multi-frequency campaign.
Finally, MU express special thanks to Mr. Y.Ishikawa 
for his continuing personal support and encouragement.




\end{document}